# Non linear behaviour of stock market volatility

Rosario Bartiromo[a]

Istituto di Struttura della Materia del CNR, via Fosso del Cavaliere 100, I-00133 Roma, Italy  and

Unita' INFM, Universita' di Roma Tre, via della Vasca Navale 84, I-00146 Roma, Italy

We exploit a continuous time random walk description of stock prices to obtain a fast and accurate evaluation of their volatility from intraday data. We show that financial markets are usefully described as open physical systems. Indeed we find that the process determining market volatility is not stationary while the market response to external volatility shocks stays constant over the time period of more than two years covered by our experimental data. Furthermore the autocorrelation function of volatility increments yields a value of about −0.4 at one-day time lag that is nearly equal for all stocks we analyze. Conditioning the evaluation of the autocorrelation function, we show that the market response is non-linear and strongly stabilizing when external shocks push for higher volatility. This market behavior can be explained by the action of participants with different time horizon.

89.65.Gh, 05.40.Fb, 05.45.Tp, 89.75.Da

[a] e-mail: bartiromo@fis.uniroma3.it

Modern financial markets are supposed to serve two main purposes. Broadly speaking, first they should be able to integrate all available information to produce a fair pricing of traded assets and secondly they should minimize the risk of holding financial activities by distributing it among all market participants.

While there is widespread agreement among economists that markets behave almost ideally with respect to their first task[1], it has been remarked that the level of price fluctuations seems too high to be justified by economic fundamentals[2]. Therefore there are concerns that markets are not yet optimized with respect to their second task.

In this paper we show how a physical approach which aims at describing financial markets as driven systems can significantly advance our understanding of their behaviour, esthablish new experimental results and suggest ways to improve their functioning. Our main result is the discovery of a significant and universal delayed response to external shocks which is non linear and stabilizing.

Volatility represents a measure of market fluctuations and is given by the relative standard deviation of daily price increments on a given time frame, usually one year. However this definition does not lead to a viable quantitative procedure to actually evaluate this important parameter since volatility is itself time dependent, often with a short time constant. Furthermore non-gaussian distributions of daily price changes are observed for many kinds of assets thus complicating data analysis[3].

For these reasons in econometrics volatility is not a measured parameter but rather it is estimated from data series of daily closing prices by means of models like for example the very popular ARCH[4] and GARCH[5]. As a result information on its behavior on short time scales is scarce. However the use of closing price is very restrictive since financial markets produce a large wealth of additional information by registering every transaction and every quotation of traded assets. Their use in volatility studies has been so far very limited because of the lack of a validated algorithm to extract the required information.

To overcome this difficulty, in a recent paper[6] we have shown that, at the single transaction level, stock prices are well described by a continuous time random walk process. This can be obtained provided a proper physical description of the system is adopted that requires to characterize the stock price by the couple of its best ask and best bid values. Then the system makes a transition only when this couple changes, the step amplitude δ being given by the corresponding price variation. We have shown that the propagator of this random walk is gaussian with a diffusion coefficient given by $D = \frac{<\delta^2>}{2<\tau>}$ where $<\delta^2>$ is the second moment of the step size distribution and $<\tau>$ is first moment of the distribution of waiting time τ between two consecutive transitions. Since markets fluctuate, the value of D depends on the time period used to evaluate $<\tau>$ and $<\delta^2>$. In this paper we will average over one day both to obtain a good precision, since many transitions are usually observed in one day for liquid stocks, and to average over the daily pattern usually observed in market activity[7]. In this way we first obtain for every market day an independent measure of D, then we exploit the properties of the random walk propagator to obtain the daily standard deviation of price increments as $\sqrt{2D \, \Delta t}$, with Δt equal to one day, and finally we translate this to a value $\sigma_t$ of annualized volatility by conventionally assuming 250 trading days in one year.

To illustrate the outcome of this approach we show in fig. 1 the results obtained for the FIB index, a future contract on the MIB, a stock average index of the Milan stock exchange. These data, as all the others we will discuss in this paper, refer to about 600 trading days from January 2001 to May 2003. Very prominent are the volatility peak after the events of 11 September 2001 and the period of market turbulence related to the emergence of financial scandals at Wall Street during the second half of 2002.

Visual inspection of fig. 1 readily shows that our data reproduce two well-known features of volatility: high values tend to cluster together (volatility clustering) while they slowly decay toward the mean value (mean reversion).

Another well-documented feature of stock volatility is the slow decay of its autocorrelation function as a function of time lag. This is represented in the upper panel of fig. 2 for our data. The continuous line refers to the whole period and indeed displays the expected behavior with significant correlation for time lags exceeding 20 days. However our data also show that the correlation function is not constant over the time period covered by the dataset. Indeed, if we divide the data points in two halves, each of them consisting of 300 consecutive observations, we find that the second half, see the dotted-dashed line in fig. 2, displays a faster autocorrelation decay when compared with the first period represented by the dashed line, resulting in a difference of nearly 50% at time lag of 20 days.

Our interpretation of this finding is that the process determining volatility behavior is not stationary. This should not be surprising, bearing in mind that the value of financial assets is influenced by a variety of independent factors such as geopolitical stability, the economic cycle or the monetary policy of central banks, just to mention a few. It is therefore very unlikely that such diverse factors could combine to yield a stationary behavior.

This reflects the fact that in physical terms financial markets are best described as open systems since they are influenced to a large extent by external events. In these circumstances what makes physical sense is the response of the market to an external shock. To this purpose we construct the serie of relative increments between two consecutive daily volatility values $\Delta\ln\sigma_t = \ln\sigma_t - \ln\sigma_{t-1}$ and we study its autocorrelation function. As shown in the lower panel of fig. 2, at one day time lag we find a significant anti-correlation $\rho_1 = -0.4$, quickly decaying below the noise level. This feature however is stable as deduced by dividing the dataset in the manner described above. This result is documented in the lower panel of fig. 2 by comparing the different lines and gives evidence that the market responds to a volatility shock always in the same way, at least for the time period we analyze.

To gain further insight in the market behavior we extended this analysis to 31 of the most liquid stocks traded in the Milan exchange. The open dots in fig. 3 represent the value of $\rho_1$ for these stocks and for the FIB index. Error bars are evaluated as the standard deviation of points with lags in the range 80-100 days. The plot clearly shows that, although the stocks are widely different among themselves with respect to liquidity and market sector, nevertheless $\rho_1$ is very similar for all data series. This is a remarkable feature, undisclosed so far in the econometric literature as far as we know, which calls for an explanation.

We first consider the possibility that all stocks are affected at any given time similarly by external factors. This would imply that their volatilities are tightly cross-correlated. To quantify this effect we studied the correlation of each stock with the FIB index, which can be taken as a good indicator of the market as a whole. The zero lag cross correlation is represented as full dots in fig. 3. A significant correlation is observed only for about one third of the stocks, the others showing a value comparable with the noise level. In any case the measured cross correlation is lower than 0.3, a value too low to explain the constancy of $\rho_1$.

In the following of this letter we will exploit the similarity between different stocks to average their autocorrelation functions and obtain a better signal to noise ratio. This average is plotted in the upper panel of fig 4 as a continuous line and shows that a significant correlation is obtained in this way also at two days lag.

In the econometric literature the use of linear stochastic processes to describe observed correlations is widespread. Adopting a similar approach, it is easily shown that a second order moving average process MA(2) can be fitted to our data. In this approximation the observed volatility variation $\Delta\sigma_t$ would be described as $\Delta\sigma_t = a_t + \varphi_1 a_{t-1} + \varphi_2 a_{t-2}$, where the $a_t$ are uncorrelated external shocks. The coefficients turn out to be $\phi_1 \sim -0.56$ and $\phi_2 \sim -0.08$ indicating that on average the market is able to recoup more than 50% of a shock with one day delay while a further 8% is absorbed the following day. A similar conclusion can be reached if a mixed autoregressive and moving average process is adopted such as for example an ARMA(1,1), widely popular in econometric, while a pure autoregressive approach yields poor results indicating once again that the volatility dynamics is dominated by external events.

However in a linear process both negative and positive shocks are dealt with in the same way. Therefore in this case, although the market would be able to reduce volatility after a positive shock, it would also increase it once a negative shock takes place. A deeper analysis of the available experimental data shows that markets do not respond linearly to volatility shocks. Conditioning the evaluation of the correlation function helps to obtain more specific information. For example if we limit our analysis to data whose amplitude of the volatility change is lower than 10% in absolute value we obtain the dotted-dashed curve in the upper panel of fig. 4. It shows that the lagged response is very much reduced in this case. We can therefore assume that the market can deal with small shocks within the same day they show up.

On the contrary if we analyze the behavior when the absolute variation is larger than 30% we observe the dashed line in the upper panel of fig. 4. In this case the lagged response is enhanced by more than 10% with respect to the full line and is about one order of magnitude larger than in the small amplitude case.

Further information is obtained when we distinguish the sign of the variations. If the volatility increases by more than 30% we observe the full curve in the lower panel of fig. 4 while for a reduction of more than 30% we obtaine the dashed line. This plot is very telling. Indeed it shows that when a shock occurs which pushes volatility up by more than 30%, the market on average is able to reduce it in the following day by a large portion. On the contrary large reduction in volatility is mostly produced in response to a comparable increase occurring on the previous day and does not lead to any significant reaction the following day.

These observations can be understood if one keeps in mind that the market is populated by a variety of investors with different time horizons[8]. However most of the wealth stays with institutional investors such as mutual or pension funds that usually update their market position only on a daily or even weekly base. Only a fraction of investors take action during the intraday session, notably among them are market makers and hedge funds. When a volatility shock takes place there is an increasing demand to sell or to buy stocks, depending on the kind of shock in action. If it is too large market participants cannot satisfy this demand immediately. At the end of the session investors with a longer time horizon take note of the situation and the following day they may decide to update their market position. This increases market participation and leads to a lower volatility. Moreover, since institutional investors tend to hold assets in proportion of their capitalization, their impact on volatility will be similar for each stock.

In conclusion we have shown that a proper physical description of stock price dynamics yields a fast and accurate evaluation of market volatility by exploiting intraday data. In this way we could characterize the market as an open physical system and uncover that there is a significant delayed response to volatility shocks. This response is non linear and stabilizing since it only reduces volatility and is likely due to investors updating their market position after the closing of the daily session. Our data also show that most of the market participants respond within a few days to external events. We impute this to the increasing use of information technology tools by investors. It is widely held that the advent of these tools is largely responsible for the observed trend of market volatility reduction. Our findings suggest that markets can still reduce their volatility by a significant amount by bringing forward the delayed response we observe.

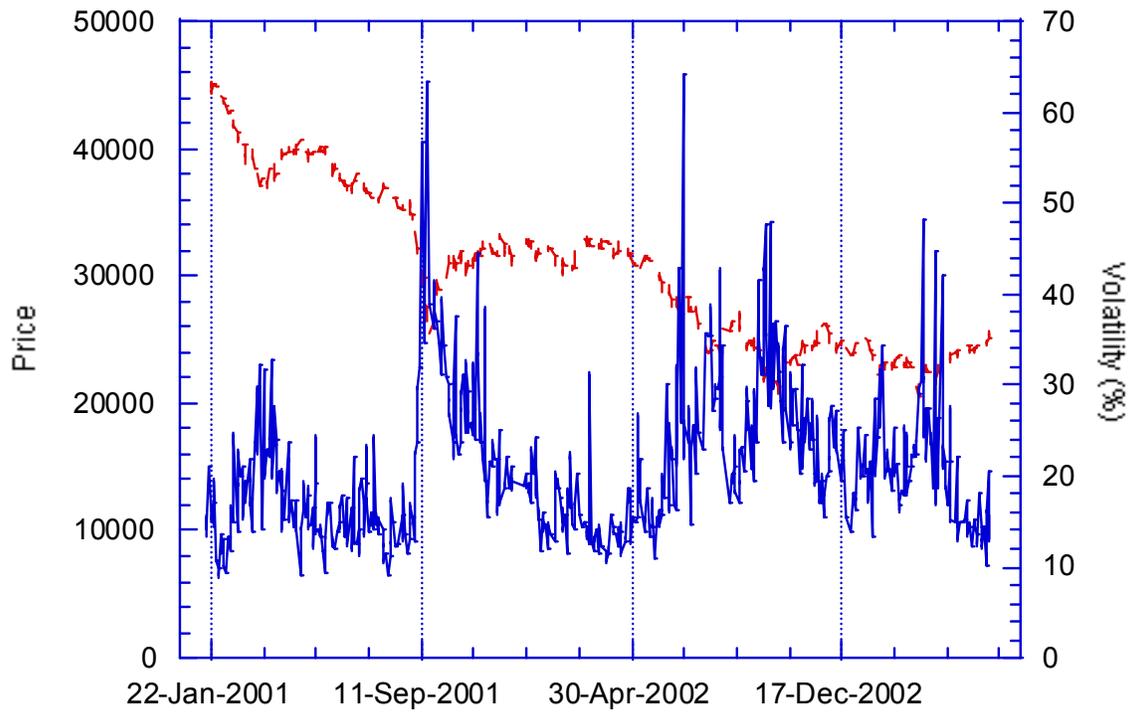

**Fig. 1.** Experimental data for the volatility of the FIB future is shown as a continuous line together with its daily closing price represented by the dashed line. The data cover the period from January 2001 to May 2003. The main peaks are due to the terrorist attack on September 11 and to the emergence of financial scandals at Wall Street in the second half of 2002. The data show both volatility clustering and mean reversion.

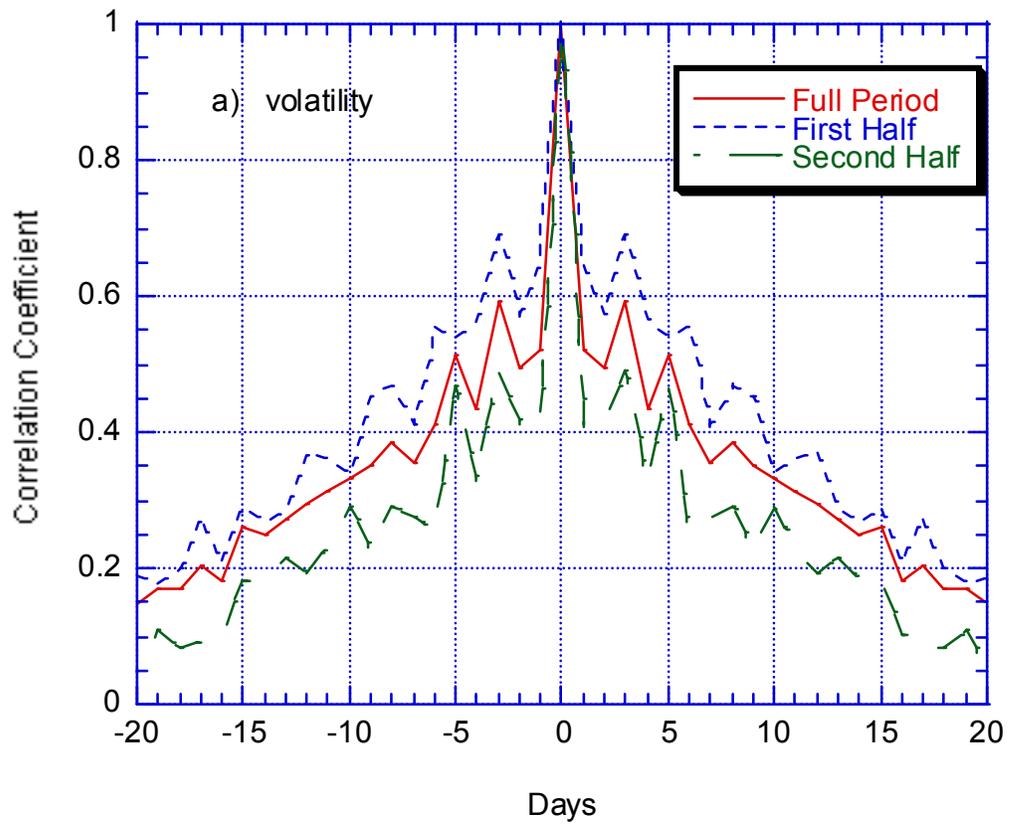

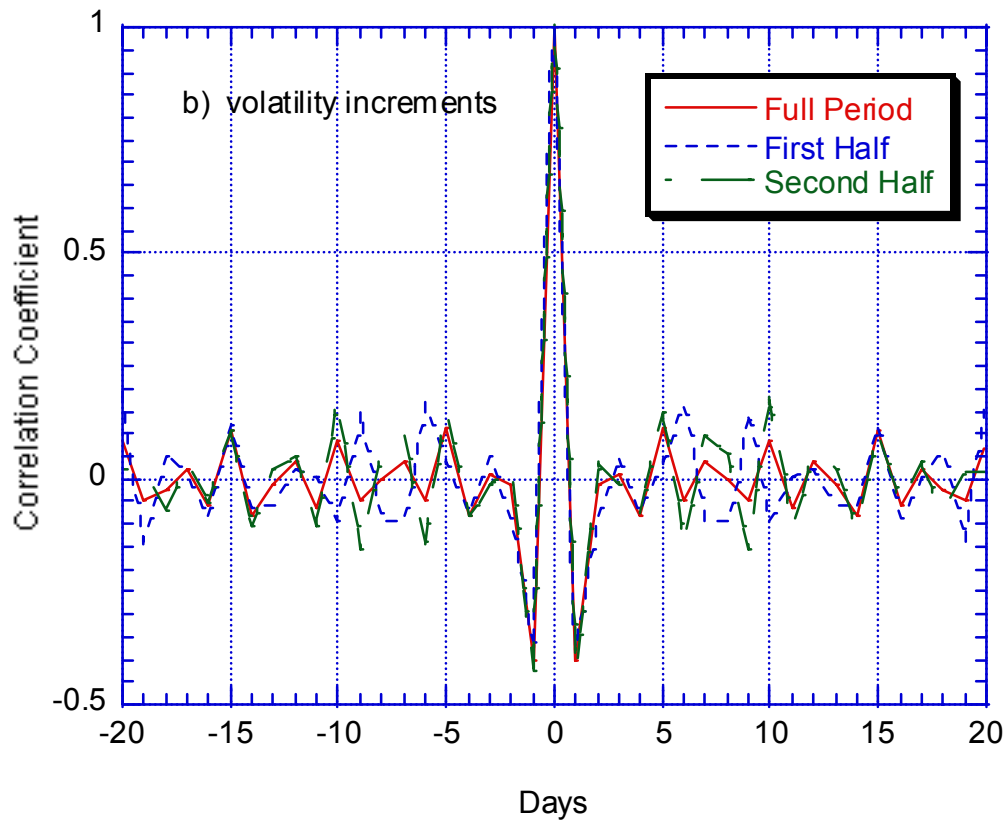

Fig. 2. (a) The autocorrelation function of FIB volatility is plotted as a continuous line and shows, as expected, a slow decay with time lag with significant correlation up to more than 20 trading days. Comparing the behavior of the first half of the dataset (dashed line) with the second half (dot dashed line) we show that autocorrelation is not stationary. (b) The series of volatility increments shows a significant value only at one-day lag and then falls below the noise level. This value is equal for the first (dashed line) and the second (dot dashed line) half of the dataset indicating a stationary mechanism for the market response to volatility shocks.

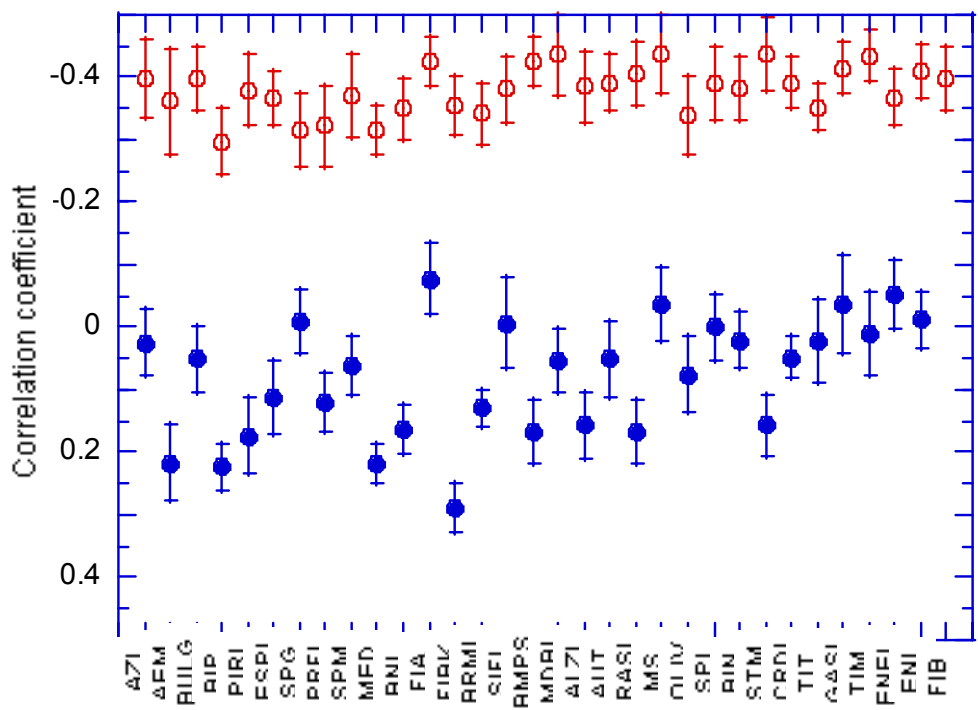

Fig. 3. The value at one-day lag of the autocorrelation function is shown as open dots for 31 stocks traded in Milan and the FIB future. Also shown with full dots is the cross correlation between each of the stocks and the FIB. On the abscissa stocks are ordered roughly with increasing capitalization. Please note that the ordinate axis is inverted.

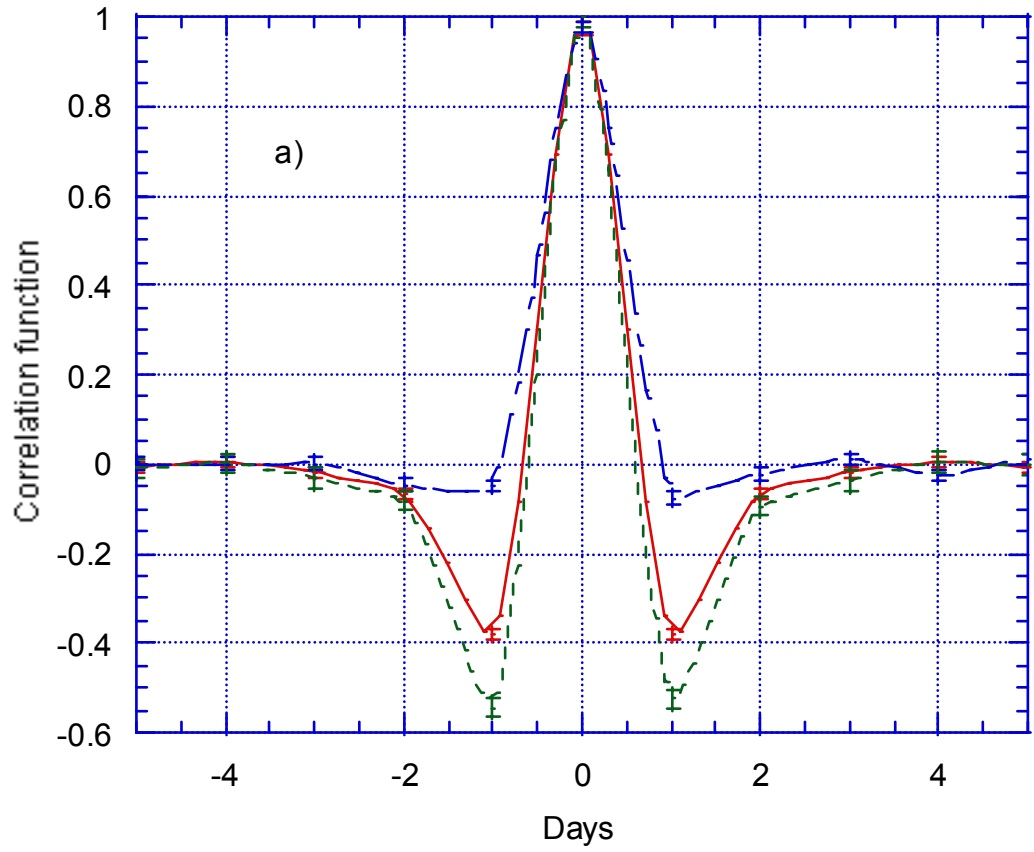

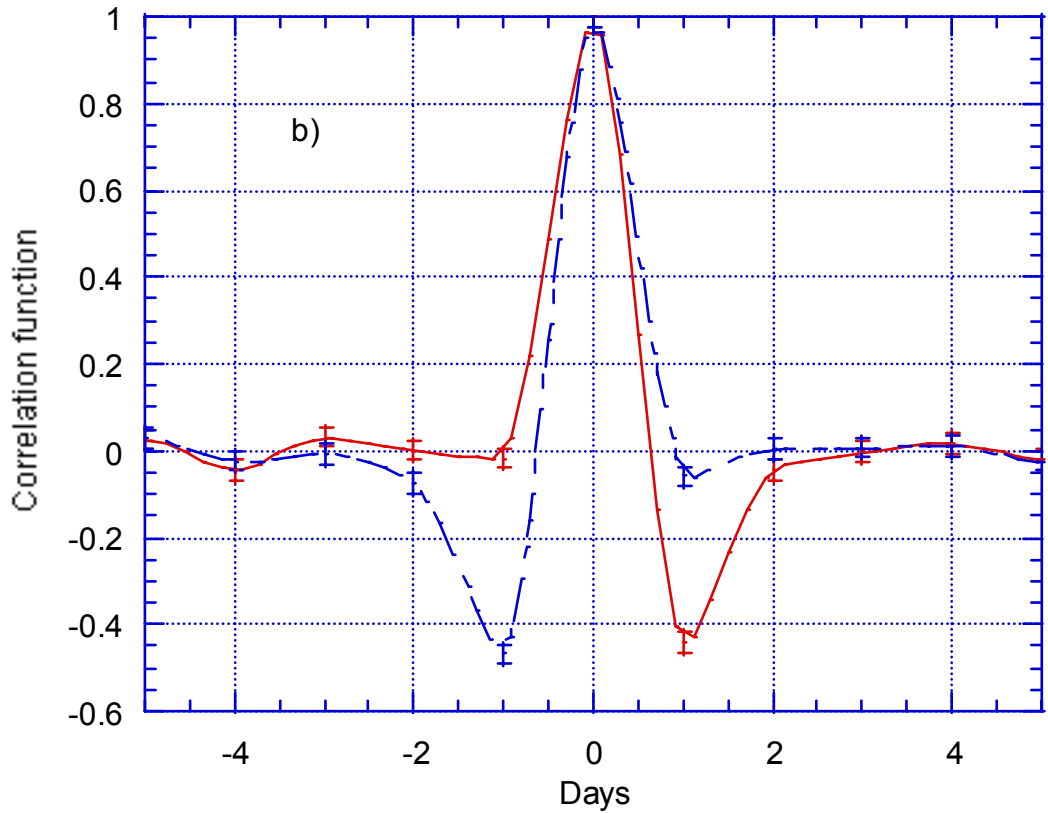

Fig. 4. (a) The autocorrelation function averaged over the 32 data series of volatility increments is shown as a continuous line. Restricting the evaluation of the autocorrelation only to data points corresponding to an increment lower than 10% in absolute value we obtain the dotted-dashed line. On the contrary data corresponding to increments larger than 30% in absolute value yield the dashed line. (b) The autocorrelation obtained for volatility increase above 30% is shown as a continuous line while the dotted-dashed line shows the behavior when volatility decreases by more than 30%.